\documentclass[12pt,english]{article}
\usepackage[T1]{fontenc}
\usepackage[utf8]{inputenc}
\usepackage{dsfont}
\usepackage{amsmath}
\usepackage{amssymb}
\usepackage{amsthm}
\usepackage{upgreek}

\usepackage{subcaption}

\usepackage{relsize}

\theoremstyle{plain}

\newtheorem*{corollary}{Corollary}

\theoremstyle{definition}

\usepackage{float}
\usepackage{algorithm}
\usepackage{algpseudocode}

\usepackage{dsfont}
\usepackage{xcolor}
\usepackage{hyperref}
\usepackage{cleveref}
\hypersetup{
    colorlinks=true,
    linkcolor=violet!100!black,
    citecolor=violet!100!black,
    urlcolor=violet!100!black
}

\usepackage{tikz}

\usetikzlibrary{positioning}
\usetikzlibrary{quantikz2}
\usepackage{graphicx}

\makeatletter
\@ifundefined{date}{}{\date{}}
\makeatletter
\newcommand{\margenes}[4]{%
\setlength{\oddsidemargin}{\dimexpr #1-\hoffset-1in}%
\setlength{\textwidth}{\dimexpr \paperwidth-#1-#2}%
\setlength{\topmargin}{\dimexpr #3-\voffset-1in}%
\setlength{\textheight}{\dimexpr \paperheight-#4-#3-\footskip-\headheight-\headsep}%
\if@twoside
  \setlength{\evensidemargin}{\dimexpr #2-\hoffset-1in}%
\fi}
\makeatother

\margenes{2cm}{2cm}{2cm}{2cm}

\crefname{section}{Sec.}{Sec.}
\crefname{equation}{Eq.}{Eqs.}
\Crefname{equation}{Eq.}{Eqs.}

\makeatother

\usepackage{babel}
\usepackage{bm}
\newcommand{\dd}{\mathrm{d}}
\newcommand{\ii}{\mathrm{i}}

\newcommand{\expect}[2][]{\langle #2\rangle_{#1}}

\algrenewcommand\algorithmicrequire{\textbf{input:}}
\algrenewcommand\algorithmicensure{\textbf{output:}}
\algrenewcommand\algorithmicindent{2.5em}

\usepackage{authblk}
\usepackage{array}
\usepackage{tabularx}

\begin{document}

\begin{center}

{\Large\bfseries Constrained Quantum Imaginary Time in Krylov Space}

\vspace{1.5em}

\begin{tabular}{cccc}
\textbf{Daniele Cuomo}$^{\star}$ & 
\textbf{Luis Cort Barrada}$^{\diamond}$ & \textbf{Micha\l{} Studzi\'nski}$^{\diamond}$ & \textbf{Robert van Leeuwen}$^{\star}$
\end{tabular}

\vspace{0.8em}

\begin{tabular}{ccc}
$^{\star}$\textit{University of Jyv\"askyl\"a, Finland} &&
$^{\diamond}$\textit{University of Gda\'{n}sk, Poland}
\end{tabular}

\vspace{0.5em}

\end{center}

\thispagestyle{empty}

\begin{abstract}
We introduce constrained quantum imaginary time evolution (\texttt{CQITE}), extending quantum imaginary time evolution (\texttt{QITE}) to constrained-search dynamics for ground-state preparation with prescribed expectation-value constraints. A central challenge in practical \texttt{QITE} is constructing a compact unitary representation of the imaginary-time evolution while maintaining accuracy and controlling the approximation error.

We resolve this challenge by deriving an exact characterization of the imaginary-time evolution from the Krylov space generated by repeated applications of the Hamiltonian to the initial state. This characterization induces a hierarchy of operator directions expressed through nested anti-commutators. Truncating the hierarchy produces a systematic static operator pool, with polynomial size in the system size at any fixed truncation order. The resulting construction provides a principled route to compact unitary approximations of imaginary-time dynamics.

Numerical tests on spin models show close agreement with exact constrained imaginary-time evolution and small constraint deviations, while a fermionic example yields energies close to the analytic constrained-search reference.
\end{abstract}



\tableofcontents

\section{Introduction}

Preparing ground states of quantum many-body systems is a central task in quantum simulation, quantum chemistry, and combinatorial optimization. Imaginary time evolution provides an effective route toward ground-state preparation by exponentially suppressing excited-state components. However, its application to large-scale quantum systems is hindered by two fundamental obstacles: the exponential growth of the many-body Hilbert space makes exact classical simulations intractable, while its intrinsically non-unitary dynamics prevents direct implementation on gate-based quantum computers \cite{McArdle2019,Yuan2019,Cuomo2026,Gluza2026,Mao2023}.

Quantum imaginary time evolution (\texttt{QITE})~\cite{Motta2020} addresses this limitation by replacing infinitesimal imaginary-time steps with locally optimal unitary transformations that reproduce the imaginary-time dynamics. This hybrid quantum--classical framework has emerged as a promising approach for ground-state preparation on gate-based quantum devices through shallow unitary updates \cite{Li2017,Yuan2019,Endo2020}.

In many applications, however, the target ground state is required to satisfy additional physical constraints, such as prescribed expectation values of conserved observables, symmetry conditions, or feasibility requirements arising in constrained optimization problems \cite{Levy1979,Lieb1983,Penz2021,Penz2025}. Constrained-search imaginary time evolution extends the imaginary-time formalism by introducing time-dependent Lagrange multipliers that enforce such constraints throughout the evolution \cite{Penz2025}. Since the resulting dynamics remain intrinsically non-unitary, their direct realization on gate-based quantum hardware remains an open challenge.

In this work we introduce the  \emph{constrained quantum imaginary time evolution} (\texttt{CQITE}) algorithm, extending the standard \texttt{QITE} algorithm to constrained-search dynamics.

The implementation of both \texttt{QITE} and \texttt{CQITE} requires the definition of a unitary operator that mimics the imaginary-time dynamics. In practice, this requires a finite support. Although the full Pauli support can represent the exact dynamics, its size grows exponentially with the system size. It is therefore not evident how to construct a compact operator support without sacrificing accuracy. Existing approaches can construct the support adaptively, but they do not directly characterize the operator directions generated by the imaginary-time evolution \cite{Grimsley2019,Yordanov2021}.

We address this problem by deriving an exact characterization of the imaginary-time evolution based on a Krylov space generated from the initial state through repeated anti-commutator actions of the Hamiltonian. 

Unlike conventional quantum Krylov methods, which construct Krylov spaces from Hamiltonian powers or real-time evolution \cite{Cortes2022,Stair2020,Kirby2023,Zhang2024}, the Krylov hierarchy derived here follows directly from the operator structure of imaginary-time evolution.
Truncating this hierarchy yields a static operator whose size scales polynomially with the system size for fixed truncation order. Increasing the truncation order enlarges the accessible space and systematically approaches the full Krylov space. The resulting construction provides an efficient alternative to variational operator pools that can be applied to both unconstrained and constrained quantum imaginary time evolution.

The resulting framework therefore provides two complementary advances. First, it extends quantum imaginary-time evolution to constrained-search dynamics, yielding a practical algorithm for constrained ground-state preparation. Second, it establishes an exact Krylov characterization of the operator directions generated by imaginary-time evolution.

The remainder of the paper is organized as follows. \cref{sec:Constrained-search-in} reviews constrained imaginary-time evolution and the constrained-search formalism underlying the proposed approach. \cref{sec:cqite-quantum-computer} derives the \texttt{CQITE} algorithm. \cref{sec:poolde} develops the Krylov hierarchy. \cref{sec:experiments} presents numerical validation. Technical details concerning the preparation of the constrained reference state, the polynomial approximation of the density operator and the preservation of the Krylov support under constraints are collected in the appendices.

\begin{algorithm}[H]
\footnotesize
\caption{\texttt{Constrained \textcolor{violet}{Quantum} Imaginary Time Evolution}}
\label{alg:CQITE}
\begin{algorithmic}[1]

\Require Hamiltonian $H$, constraint $\mathbf C$, target $\mathbf b$\Comment{Pauli data are efficiently encoded}

\vspace{0.6em}

\State \makebox[1.5cm][l]{\textbf{hierarchy}}\quad 
 $\mathbf{P}
\ \gets\ \operatorname{vec} \bigcup_{j}\operatorname{supp}H^j$
\Comment{\textbf{\cref{sec:poolde}}}

\vspace{0.6em}
\State \makebox[1.7cm][l]{\textbf{reference}} $|\varphi(0)\rangle \gets \textbf{any}\ |\varphi\rangle \textbf{ s.t. } \langle\mathbf C\rangle=\mathbf b , \qquad \tau \gets 0$
\Comment{\textbf{\cref{sec:init}\!}}

\vspace{0.6em}

\Repeat

\vspace{0.6em}
\color{violet}
    
    \State \textbf{evolve}\Comment{Run multiple times for sampling} \[|\varphi(\tau)\rangle
    \ \gets\ 
    \mathcal{T}\!\exp\!\Big[{-\ii\!\int_0^{\tau}\!\dd\varsigma A(\varsigma)}\Big]|\varphi(0)\rangle\]

\State \textbf{estimate \cref{eq:Constraint1} and \cref{eq:PP}}
\color{black}
\vspace{0.6em}

\State \makebox[1.2cm][l]{\textbf{enforce}}\quad $
\mathbf{v}\phantom{\mathbf a} \gets\ 
-\langle\!\langle \mathbf C\mathbf C^{\mathsf T}\rangle\!\rangle^{\!\phantom{.\!}^{\!\texttt{-1}}}\langle\!\langle H\mathbf C\rangle\!\rangle
$
\Comment{\textbf{\cref{sec:Constrained-search-in}}}

\vspace{0.6em}

\State \makebox[1.2cm][l]{\textbf{project}}\quad
$\mathbf a\phantom{\mathbf{v}}
\gets\ 
- \operatorname{Re}\langle\mathbf P\mathbf P^{\mathsf T}\rangle^{\!\phantom{.\!}^{\!\texttt{-1}}}\operatorname{Im}\langle\mathbf P(H + \mathbf C^{\mathsf T}\mathbf{v})\rangle
$
\Comment{\textbf{\cref{sec:cqite-quantum-computer}}}

\vspace{0.6em}
\State \makebox[1.2cm][l]{\textbf{update}}\quad $
A\big|_{(\tau,\tau + \updelta]}
\ \gets\ \mathbf{a}^{\mathsf{T}}\mathbf{P},\qquad \tau\gets \tau + \updelta$

\vspace{0.6em}

\Until{\textbf{\cref{eq:convergence}}}

\vspace{0.6em}
\State \Return $\langle H\rangle$

\end{algorithmic}
\end{algorithm}

\section{Imaginary time under constraints\label{sec:Constrained-search-in}}
Let
\begin{equation}
\mathbf C
=
(C_1,\ldots,C_m)^{\mathsf T}\in \operatorname{span}\mathcal{P}^m_w.
\end{equation}
be a vector of constraints, where each \(C_j\) is expressed as a linear combination of operators coming from the Pauli group $\mathcal{P}_w = \{P_1\otimes\cdots\otimes P_w\ |\ P\in \{\mathds{1},X,Y,Z\}\}$.


%

Given a target vector $\mathbf b\in\mathbb R^m$, we consider the constrained optimization problem
\begin{equation}
\begin{aligned}
\min_{|\varphi\rangle}\quad &
\langle H\rangle \\
\mathrm{s.t.}\quad &
\mathbf \langle \mathbf{C}\rangle=\mathbf b,
\end{aligned}
\label{eq:general_constraint_problem}
\end{equation}
with $H$ being a linear combination of $n$ operators in $\mathcal{P}_w$.


Following the constrained-search formalism
\cite{Levy1979,Lieb1983}
recently extended to imaginary time
\cite{Penz2025}, the optimization method is carried out by solving the differential equation
\begin{equation}
-\partial_\tau |\varphi(\tau)\rangle
=
\bigl(G(\tau)-\langle G(\tau)\rangle\bigr)|\varphi(\tau)\rangle,
\qquad
G(\tau)
=
H+\mathbf{v}(\tau)^{\mathsf T}\mathbf C,
\label{eq:SchrodingerImaginary}
\end{equation}
where $\mathbf{v}
 \in \mathbb{R}^m$ is the vector of real time-dependent Lagrange multipliers enforcing the constraints.
 In the limit $\tau \rightarrow \infty$ the term $\mathbf{v}^{\mathsf T}\mathbf C$ represents an external potential which added to the Hamiltonian ensures that the ground state of the system in the presence of this extra potential $\mathbf{v}$ has the prescribed expectation values. It can thus be seen as control potential that can adjust ground states so that they have desired expectation values. A good physical example is to tune an external magnetic field for a spin-lattice until a desired spin-magnetization expectation value is achieved in the ground state. 

Preservation of the constraints requires
\begin{equation}
\label{eq:dC}
\frac{\dd}{\dd\tau}
\langle \mathbf C \rangle
=
0.
\end{equation}

Using \cref{eq:SchrodingerImaginary}, the Lagrange multipliers are determined by the real linear system
\begin{equation}
\langle\!\langle
\mathbf{C}\mathbf{C}^{\mathsf T}
\rangle\!\rangle
\mathbf{v}
=
-\langle\!\langle
H\mathbf{C}
\rangle\!\rangle,
\label{eq:Constraint1}
\end{equation}
where $\langle\!\langle\cdot\rangle\!\rangle$ denotes the symmetrized covariance.

The matrix $\langle\!\langle
\mathbf{C}\mathbf{C}^{\mathsf T}
\rangle\!\rangle$ is state-dependent and it is not guaranteed to be invertible for every state $\ket{\varphi}$. If the constraint matrix is singular, the multipliers are understood in the Moore--Penrose pseudoinverse sense. 

At convergence, the flow reaches a stationary point satisfying
\begin{equation}
\label{eq:convergence}
\bigl(G-\langle G\rangle\bigr)|\varphi\rangle=0 .
\end{equation}

This condition characterizes a critical point of the energy functional, and it can be shown that such a stationary point corresponds to an absolute minimizer under the constraints \cite{Penz2025}, provided that $\ket{\varphi(0)}$ has a non-zero overlap with it.




\section{Constrained QITE}\label{sec:cqite-quantum-computer}

We solve the imaginary time equation \cref{eq:SchrodingerImaginary} by assuming that the generator is piecewise constant in $\updelta$.

The step generator  $G$ is built from the incoming state, according to \cref{eq:Constraint1}.

For an incoming state $|\varphi(\tau - \updelta)\rangle$, the exact normalized constrained imaginary time step is given by
\begin{equation}
|\varphi_\star(\tau) \rangle=\frac{e^{- \updelta G}}{\sqrt{\langle e^{-2 \updelta G }\rangle}}|\varphi(\tau-\updelta)\rangle.
\label{eq:exact-step}
\end{equation}

The exact step \cref{eq:exact-step} is non-unitary. Since both $|\varphi(\tau-\updelta) \rangle$ and $|\varphi_\star(\tau) \rangle$ are normalized, there exists at least one unitary operator $U_\star$ such that
\begin{equation}
|\varphi_\star(\tau) \rangle= U_\star|\varphi(\tau- \updelta) \rangle,
\end{equation}
under the piecewise constant assumption.

The \texttt{CQITE} step seeks a structured Hermitian operator $A$ such that
\begin{equation}
\ket{\varphi(\tau)} = e^{- \ii \updelta A}|\varphi(\tau-\updelta)\rangle
\end{equation}
approximates the exact step \cref{eq:exact-step} locally in $\updelta$.

\subsection{Operator expansion and variational equations}
\label{subsec:operator-expansion}

To construct a practical approximation to the normalized imaginary time
evolution, we expand the unitary generator in a restricted operator pool $\mathbf P =
(P_1,\ldots,P_{w})^{\mathsf{T}}$,
and coefficients $\mathbf a
=
(a_1,\ldots,a_{w})^{\mathsf{T}}$, for some $w \geq 1$.

The effective generator is written as
\begin{equation}
A
= \mathbf a^{\mathsf T}\mathbf P,
\label{eq:AExpansion}
\end{equation}
where the coefficients are determined variationally \cite{Li2017,Yuan2019,Endo2020}.

Expanding the normalized imaginary time step to first order in $\updelta$ gives
\begin{equation}
|\varphi_\star(\tau)\rangle
=
|\varphi(\tau - \updelta)\rangle -
\updelta\bigl(
G
-
\expect[]{G}
\bigr)
|\varphi(\tau - \updelta)\rangle + \mathcal{O}(\updelta^2),
\end{equation}
Expanding the variational unitary ansatz to first order in $\updelta$
yields
\begin{equation}
|\varphi(\tau)\rangle
= |\varphi(\tau-\updelta)\rangle
-\ii \updelta A
|\varphi(\tau-\updelta)\rangle + \mathcal{O}(\updelta^2).
\label{eq:apx_step}
\end{equation}
The coefficients of $A$ are obtained by requiring stationarity of the
squared distance between the exact tangent vector and its variational
approximation~\cite{McLachlan1964,Yuan2019,McArdle2019},
\begin{equation}
\nabla_{\mathbf a}
\langle
\varphi_\star(\tau)-\varphi(\tau)
\,|\,
\varphi_\star(\tau)-\varphi(\tau)
\rangle
=
 0.
\end{equation}
Matching the first-order terms in the two expansions projects the exact
imaginary time update onto the variational subspace spanned by
$\nabla_{\mathbf a}|\varphi(\tau-\updelta)\rangle$ and yields the linear system
\begin{equation}
\label{eq:PP}
\operatorname{Re}\langle \mathbf P\mathbf P^{\mathsf T}\rangle
\mathbf a
=
-\operatorname{Im}\langle \mathbf PG\rangle.
\end{equation}




\subsection{State-dependent occupation number preservation}
\label{sec:sq}
A large class of Hamiltonians of physical interest is that of interacting fermions, given in second quantization notation by the general form
\begin{equation}
    H = \sum_{p,q} h_{pq}\, c_p^\dagger c_q^{\phantom{\dagger}} + \sum_{p,q,s,r} v_{pqsr} \, c_p^\dagger c_q^\dagger c_s^{\phantom{\dagger}} c_r^{\phantom{\dagger}}
\end{equation}
where $c_p^\dagger$ and $c_p$ are fermionic creation an annihilation operators. This class of Hamiltonians includes the Hubbard model or the general form of a molecular Hamiltonian in quantum chemistry. An important class of constraints is that represented by one-body operators of the form
\begin{equation}
    C_j = \sum_{p,q} D^{(j)}_{pq} \, c_p^\dagger c_q^{\phantom{\dagger}}
\end{equation}
for Hermitian matrices $D^{(j)}_{pq}$. As an example of a set of commuting constraints we can take the site occupations $n_j=c_j^\dagger c_j$ of the Hubbard model or of lattice density-functional theory.
Using a qubit encoding of the
fermionic Fock space, such as the Jordan--Wigner transformation, the vector of site-occupation operators is represented as
\begin{equation}
    \mathbf{n}
    =
    \frac{1-\mathbf{Z}}{2},
\end{equation}
where
\begin{equation}
    \mathbf{n}
    =
    \left(
    c_1^\dagger c^{\phantom{\dagger}}_1,\ldots,c_m^\dagger c^{\phantom{\dagger}}_m
    \right)^{\mathsf{T}},
    \qquad
    \mathbf{Z}
    =
    \left(
    Z_1,\ldots,Z_m
    \right)^{\mathsf{T}} = \mathbf{C}
\end{equation}
are the vectors of occupation-number operators and constraints,
respectively, for $m$ fermionic modes.
The corresponding occupation expectation values are therefore given by
\begin{equation}
    \langle\mathbf{n}\rangle
    =
    \frac{1-\langle\mathbf{C}\rangle}{2}.
\end{equation}
Thus, the admissible target values satisfy
\begin{equation}
\label{eq:sq_bounds}
    -1
    \leq
    \mathbf{b}
    \leq
    1,
\end{equation}
which is equivalent to requiring
\begin{equation}
    0
    \leq
    \langle\mathbf{n}\rangle
    \leq
    1.
\end{equation}
The target values are chosen consistently with the desired $n$-particle
sector, i.e.
\begin{equation}
\label{eq:sq_tgt}
    \mathbf{1}^{\mathsf T}\mathbf{b}=m - 2n.
\end{equation}

The initial state must be prepared using number-conserving two-mode rotations. Since both the Hamiltonian and the constraints preserve particle number, the constrained imaginary-time evolution of \cref{eq:exact-step} remains in the same particle-number sector.

The \texttt{CQITE} unitary is then constructed state-dependently to reproduce
this constrained imaginary-time update. Crucially, particle-number
preservation is therefore required only for the action of the unitary on
the state generated by the algorithm, and not as an operator-level symmetry
of the unitary itself. The unitary need not preserve particle number for
arbitrary input states: it is sufficient that, for the particular input
state encountered during the evolution, it reproduces the corresponding
constrained imaginary-time state, which belongs to the same particle-number
sector. Particle-number preservation thus follows from the appropriately
initialized state and the constrained imaginary-time evolution.
However, the approximating nature of the finite operator pool may
result in particle-number leakage in the practical implementation.

\section{Static polynomial Krylov spaces}
\label{sec:poolde}

The \texttt{CQITE} formulation derived in the previous section is exact provided that the unitary generator is expanded in the complete Pauli operator basis. In practice, however, such an expansion is computationally prohibitive, since the number of Pauli operators grows exponentially with the system size. The central challenge is therefore to identify a compact operator pool that is sufficiently expressive to reproduce the (constrained) imaginary-time dynamics while scaling efficiently with system size.

The structure of the exact \texttt{QITE} generator provides a systematic route to this construction. Rather than selecting the operator pool heuristically, we derive it directly from the operator content of the exact imaginary-time evolution. We show that the relevant operator directions form a hierarchy of Krylov spaces generated by repeated actions of the Hamiltonian on the operator space. Truncating this hierarchy at a fixed order yields a static operator pool whose size grows polynomially with the system size.

The Krylov sequence provides a systematic characterization of the energy-driven directions accessible from the reference state. We therefore derive the corresponding operator hierarchy for the Hamiltonian alone, for simplicity. The resulting operator pools can then be employed in the constrained setting, where they perform well numerically, as demonstrated in \cref{sec:experiments}.

Our proposal is closely related to the broader idea of quantum Krylov subspace methods, which have recently emerged as an alternative paradigm for quantum eigensolvers based on Hamiltonian powers or real-time evolution \cite{Stair2020,Cortes2022,Kirby2023,Zhang2024,Kirby2026Observation}. Here, however, the Krylov hierarchy is not introduced as an independent variational ansatz, but follows directly from the nested anti-commutator structure of the exact imaginary-time propagator. Truncating this hierarchy therefore provides a systematic sequence of approximations with increasing expressive power.

\subsection{A unitary representation for imaginary time}
\label{sec:Adef}

Let us define the Hermitian operator
\begin{equation}
{A}(\tau)
=
-\ii\left[{G},\Phi(\tau)\right]
=
{A}^\dagger(\tau),
\label{eq:A_definition}
\end{equation}
with $\Phi(\tau) = \ket{\varphi(\tau)}\bra{\varphi(\tau)}$.

Applying $A$ to the state gives
\begin{equation}
\label{eq:A_action}
{A}(\tau)|\varphi(\tau)\rangle=-\ii
\big(
{G}\Phi(\tau)
-
\Phi(\tau)G
\big)
|\varphi(\tau)\rangle
=
-\ii\left(G-\langle G \rangle\right)|\varphi(\tau)\rangle.
\end{equation}
Therefore,
\begin{equation}
\ii\partial_\tau |\varphi(\tau)\rangle
=
A(\tau)|\varphi(\tau)\rangle,
\label{eq:schrodinger_like}
\end{equation}

which has the form of a Schr\"{o}dinger equation. A key observation is that $A(\tau)$ depends on the very state on which it acts, but we will suppress this dependence in the notation for clarity of presentation. Moreover, even if $G$ is a one-body or two-body operator, $A$ is generally an $n$-body operator.

The Schr\"{o}dinger equation can be rewritten in integral form as
\begin{equation}
|\varphi(\tau)\rangle
=
|\varphi(0)\rangle
-
\ii
\int_0^\tau
\dd \varsigma\,
A(\varsigma)|\varphi(\varsigma)\rangle.
\label{eq:integral_form}
\end{equation}

This naturally suggests the time-stepping scheme
\begin{equation}
|\varphi(\tau+\updelta)\rangle =
|\varphi(\tau)\rangle
-
\ii
\int_\tau^{\tau+\updelta}
\dd \varsigma\,
A(\varsigma)|\varphi(\varsigma)\rangle
\approx
e^{-\ii\updelta A(\tau)}
|\varphi(\tau)\rangle
\label{eq:timestep}
\end{equation}
for sufficiently small $\updelta$.

For $\tau\rightarrow\infty$, this evolution converges to the ground state provided the initial state has a nonzero overlap with the ground state.

\subsection{Krylov representation of the evolution}
\label{sec:krylov}
For simplicity, we restrict the analysis to the Hamiltonian $H$, for which we derive the exact propagator. 

The \texttt{QITE} propagator satisfies
\begin{equation}
\ket{\varphi(\tau+\updelta)}
\approx e^{-\ii\updelta A(\tau)}
|\varphi(\tau)\rangle = 
e^{-\updelta[H,\Phi(\tau)]}
\ket{\varphi(\tau)}.
\label{eq:step_commutator}
\end{equation}

To characterize the operator content of the propagator, we express the density operator directly
from the normalized imaginary time evolution. The solution of
\begin{equation}
-\partial_\tau\ket{\varphi(\tau)}
=
(H-\langle H\rangle)\ket{\varphi(\tau)}
\end{equation}
is
\begin{equation}
\ket{\varphi(\tau)}
=
\frac{e^{-\tau H}\ket{\varphi(0)}}
{\sqrt{\langle\varphi(0)|e^{-2\tau H}|\varphi(0)\rangle}}.
\end{equation}

The associated density operator takes the form
\begin{equation}
\Phi(\tau)
=
\frac{
e^{-\tau H}
\Phi(0)
e^{-\tau H}
}{
\operatorname{Tr}
\!\left(
e^{-2\tau H}\Phi(0)
\right)
}.
\label{eq:phi_tau}
\end{equation}

Since the denominator is a scalar, it factors out of the commutator:
\begin{equation}
[H,\Phi(\tau)]
=
\frac{
e^{-\tau H}
[H,\Phi(0)]
e^{-\tau H}
}{
\operatorname{Tr}
\!\left(
e^{-2\tau H}\Phi(0)
\right)
} = \ii \frac{
e^{-\tau H}
A(0)
e^{-\tau H}
}{
\operatorname{Tr}
\!\left(
e^{-2\tau H}\Phi(0)
\right)
}.
\label{eq:comm}
\end{equation}

The normalization factor only rescales the generator and does not modify its operator support.
Consequently, the operator directions are entirely determined by
\begin{equation}
A(\tau) \propto e^{-\tau H}
A(0)
e^{-\tau H}.
\end{equation}

The imaginary-time evolution satisfies
\begin{equation}
e^{-\tau H}A(0) e^{-\tau H}
=
e^{-\tau\operatorname{ac}_H}A(0),
\end{equation}
where $\operatorname{ac}_H$ denotes the anti-commutator operator $\operatorname{ac}_{H} A = HA+AH$.

Expanding the exponential,
\begin{equation}
e^{-\tau H}
A(0)
e^{-\tau H}
=
\sum_{j}^{\infty}
\frac{(-\tau)^j}{j!}
\operatorname{ac}_H^{\,j}
A(0).
\label{eq:ac_series}
\end{equation}

\begin{corollary}
    The operator directions are spanned by the Krylov subspace \cite{Lanczos1950,Park1986}
\begin{equation}
\mathcal K(\operatorname{ac}_H,A(0))
=
\operatorname{span}
\left\{
\operatorname{ac}_H^{\,j}A(0)
\right\}_{\!j}, \quad \dim\mathcal{K} < \infty
\end{equation}
\end{corollary}




The operator pool is therefore constructed from the Pauli
support of the Krylov sequence.
This choice is directly motivated by the structure of the exact
imaginary time evolution and provides a systematic hierarchy of
approximations. Increasing the truncation order enlarges the accessible
variational manifold and converges toward the full Krylov space generated
by the anticommutator superoperator.


The resulting pool provides a set of operators that
captures the dominant directions explored by the exact dynamics while
remaining independent of the instantaneous values of the variational
coefficients.

\subsection{Polynomial construction of the pool}
The construction above defines the Krylov pool through the operator
sequence $\{\operatorname{ac}_H^{\,j} A(0)\}_j$, but a direct Pauli expansion
of $A(0)$ would still require an explicit Pauli representation of
$\Phi(0)$. The dyadic structure of the initial state allows us to avoid
this expansion altogether and construct a polynomial-size superset of
the same Krylov directions directly from powers of $H$. 

The operator Krylov sequence action satisfies
\begin{equation}
    \operatorname{ac}_H^{\;j} A(0)\ket{\varphi(0)}
    \in
    \operatorname{span}\big\{\ket{\varphi(0)}, H\ket{\varphi(0)}, \dots, H^{j+1}\ket{\varphi(0)}\big\},
\end{equation}

This relation allows the Pauli pool to be constructed without expanding $\Phi(0)$. For a Hamiltonian composed by $n$ Pauli operators, each product is itself Pauli up to a
phase. The number of distinct Pauli strings in the support of $H^j$
is bounded by
\begin{equation}
\left|\operatorname{supp} H^j\right|
    \leq n^j.
\end{equation}

We therefore define the Pauli-support vector
\begin{equation}
\label{eq:P}
    \mathbf{P}
    =
    \operatorname{vec} \bigcup_{j}^c\operatorname{supp}H^j,
\end{equation}
with $\operatorname{vec}\{P_1,\dots, P_w\} = (P_1,\dots, P_w)^{\mathsf T}$.

The resulting pool contains all Pauli directions required by the
truncated Krylov construction.

Importantly, the imaginary-time parameter $\tau$ does not enter the
construction of the Pauli pool.
Hence, $\tau$ determines only the coefficients multiplying the Krylov
directions, while the Pauli pool itself is determined solely by the
Krylov order $c$ and the Hamiltonian. Therefore, the pool can be constructed once and subsequently used for
all values of $\tau$. 

The size of this pool is bounded by $\mathcal{O}(n^{c})$.
No Pauli expansion of $\Phi(0)$ is required.

\subsection{Norm and truncation bounds}

Using the Krylov expansion of the exact generator, the contribution associated with the
first \(c\) Krylov operators is contained in $\mathbf{P}$ of \cref{eq:P}, which is a vector of Pauli operators
\begin{equation}
    \mathbf{P} = (P_1,\dots,P_{\mathcal{O}(n^c)})^{\mathsf T}.
\end{equation}

Assuming that
the variational coefficients \(\mathbf{a}\) are obtained exactly, the remaining
error is therefore bounded by the contribution of the Krylov tail. Hence,
\begin{equation}
\begin{aligned}
\left\|
(A(\tau)-\mathbf{a}^{\mathsf{T}}\mathbf{P})|\varphi(\tau)\rangle
\right\|_{2}
&\lesssim
\left\|
\sum_{j>c}
\frac{(-\tau)^j}{j!}
\operatorname{ac}_H^{\;j} A(0)
|\varphi(\tau)\rangle
\right\|_{2} .
\end{aligned}
\end{equation}

The state dependence can be removed from the upper bound by using
\begin{equation}
\| \operatorname{ac}_H^{\;j} A(0)|\varphi(\tau)\rangle\|_{2}\leq\|\operatorname{ac}_H^{\;j} A(0)\|\leq 2\|H^{j+1}\|.
\end{equation}
Thus,
\begin{equation}
\left\|
(A(\tau)-\mathbf{a}^{\mathsf{T}}\mathbf{P})|\varphi(\tau)\rangle
\right\|_2
\lesssim
\sum_{j>c}
\frac{(\tau\|
H\|)^j}{j!}.
\end{equation}

Finally, applying the standard upper bound for the remainder of the exponential
series, gives the uniform bound
\begin{equation}
\label{eq:residual}
\left\|
(A(\tau)-\mathbf{a}^{\mathsf{T}}\mathbf{P})|\varphi(\tau)\rangle
\right\|_2
\lesssim
\frac{
(\tau\|H\|)^{c+1}
}{
(c+1)!
}
\end{equation}

Since the residual bound is monotonically increasing with \(\tau\), it is sufficient to evaluate the bound at the characteristic convergence timescale set by the spectral gap, i.e. $\tau \lesssim  \frac{1}{E_1 - E_0}$, and normalized by \(\|H\|\). Hence we obtain
\begin{equation}
\left\|
(A(\tau)-\mathbf{a}^{\mathsf{T}}\mathbf{P})|\varphi(\tau)\rangle
\right\|_2
\lesssim
\frac{
1
}{
(c+1)!\Delta^{c+1}
},
\end{equation}
with $\Delta = (E_1 - E_0)/\|H\|$, satisfying $0< \Delta \leq 2$. Hence, within the fast-convergence regime $\Delta\geq 1$ the residual exhibits a factorial decay with $c$.

For $\Delta<1$, a larger Krylov order may be required to achieve the same residual accuracy, e.g.
\begin{equation}
    c = \Theta(\Delta^{-1}),
\end{equation}
namely, $c$ is asymptotically bounded both below and above by $\Delta^{-1}$.

A favorable scenario is given by
$\Delta=\Theta(1)$, where $c$ can be kept constant and the pool
size therefore scales polynomially with $n$.

For gaps that close sufficiently rapidly, this state-independent
bound may become increasingly conservative, as it does not account
for additional structure that can reduce the required Krylov order.

The above derivation also allows us to obtain a dimensionless truncation measure bounded by unity. We normalize the Taylor tail by the total weight of the exponential series. In fact, for \(c\) sufficiently beyond the characteristic scale \(\Delta^{-1}\), the normalized tail satisfies
\begin{equation}
\label{eq:taylor} e^{-1/\Delta}\sum_{j>c}\frac{1}{j!\Delta^j}
\leq e^{-\Omega(1/\Delta)},
\end{equation}
and is therefore strongly suppressed, as $\Omega$ means asymptotic lower bound.



\section{Numerical validation}
\label{sec:experiments}
The following numerical examples illustrate the performance of Algorithm \ref{alg:CQITE} on constrained spin systems of increasing size. In all simulations, the operator pool is constructed from the Krylov expansion truncated at order $c=1$ as described in Sec.~\ref{sec:poolde}. Therefore,
\begin{equation}
    \mathbf{P} = (P_1,\dots,P_{\mathcal{O}(n)})^{\mathsf T}.
\end{equation}
The imaginary time evolution is performed using a constant time step $\updelta = 10^{-2}$.
Unless explicitly stated otherwise, these parameters are kept fixed throughout all experiments.

\subsection{Three-spin Ising model}

As a first benchmark, we consider the three-spin triangular Ising model.
The Hamiltonian contains only pairwise Ising interaction terms, while the optimization is performed under local constraints on the transverse magnetization~$\mathbf X =
(X_1,\ldots,X_m)^{\mathsf T}$. The resulting constrained optimization problem is
\begin{equation}
\begin{aligned}
\min_{|\varphi\rangle}\quad &
\sum_{(u,v)}\gamma_{uv} \langle Z_u Z_v\rangle \\
\mathrm{s.t.}\quad &
\langle \mathbf X\rangle=\mathbf b .
\end{aligned}
\label{eq:ising_constraint_problem}
\end{equation}
\begin{figure}[tbh]
    \centering
    \begin{subfigure}[b]{0.49\linewidth}
        \centering
        \includegraphics[width=\linewidth]{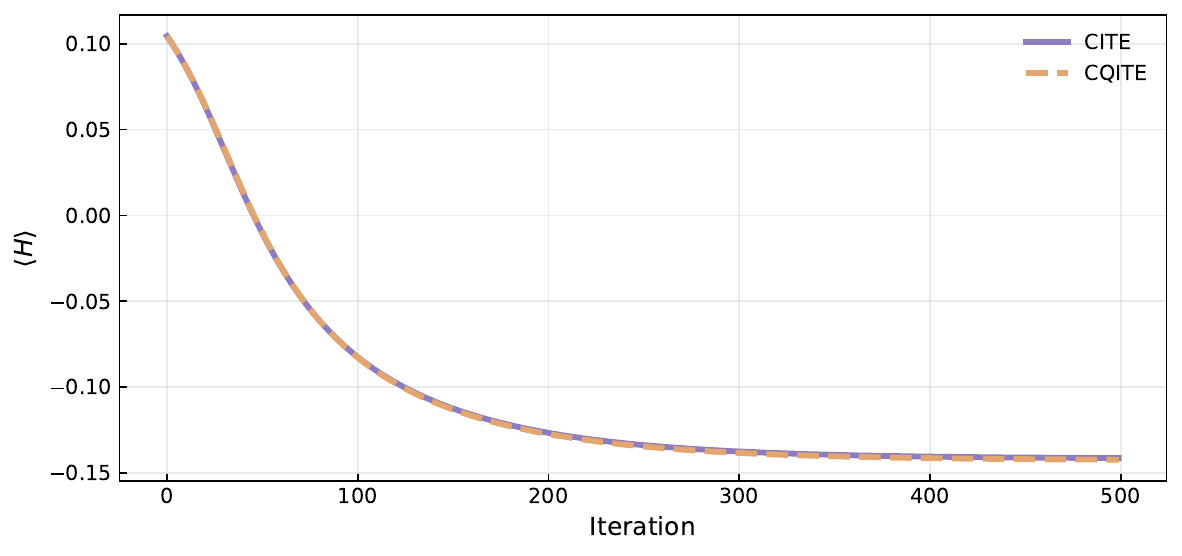}
        \caption{Energy evolution of \texttt{CQITE} and \texttt{CITE}.}
        \label{fig:zz_exp}
    \end{subfigure}
    \hfill
    \begin{subfigure}[b]{0.49\linewidth}
        \centering
        \includegraphics[width=\linewidth]{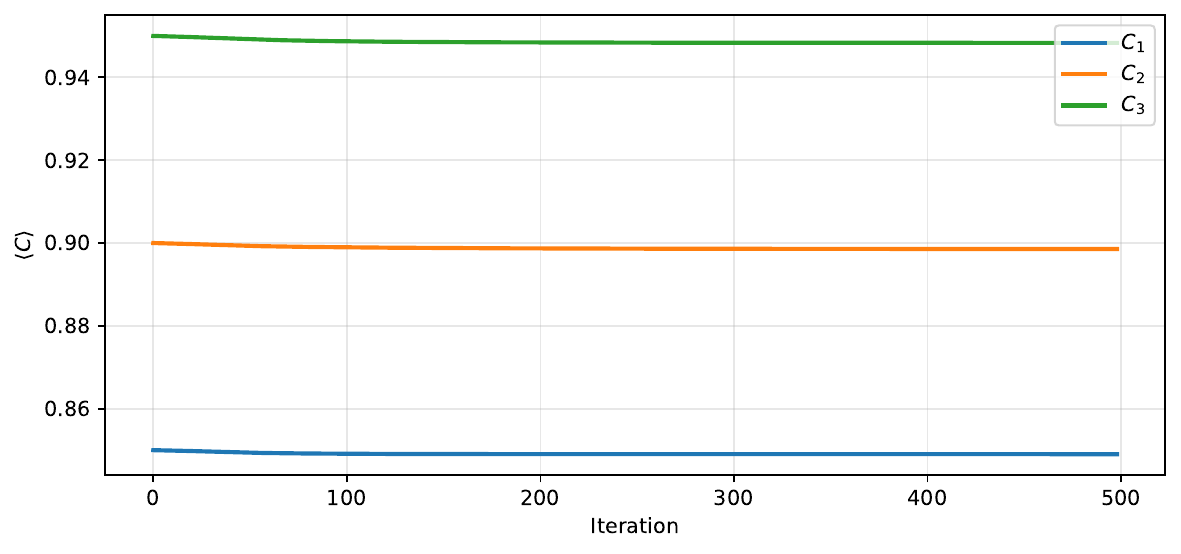}
        \caption{Constraint dynamics.}
        \label{fig:zz_con}
    \end{subfigure}    \caption{\texttt{CQITE} applied to the three-spin Ising model with transverse-magnetization constraints. (a) Energy evolution obtained with homogeneous Ising couplings $\gamma=0.2$, comparing \texttt{CQITE} with the exact \texttt{CITE}. (b) Constraint preservation $\langle \mathbf{X}\rangle$ during the evolution for target values $\bm b=(.85,.90,.95)$, showing that the prescribed constraints are accurately preserved throughout the optimization.}
    \label{fig:zz_results}
    \hrulefill
\end{figure}

Since the constraint operators do not commute with the Hamiltonian, the constrained minimum differs from the unconstrained Ising ground state, providing a simple yet nontrivial test of the \texttt{CQITE} formalism.

The system is initialized in a product state satisfying the prescribed constraint values, and both \texttt{CQITE} and the exact constrained imaginary time evolution (\texttt{CITE}) are propagated using the same imaginary time step. Figure~\ref{fig:zz_exp} compares the corresponding energy evolution. The \texttt{CQITE} trajectory closely follows the exact \texttt{CITE} dynamics throughout the evolution and converges to the same constrained minimum, confirming that the unitary approximation accurately reproduces the constrained imaginary time flow.

The evolution of the constrained expectation values is shown in Fig.~\ref{fig:zz_con}. Throughout the propagation the expectation values remain close to the prescribed targets, demonstrating that the Lagrange-multiplier formulation successfully confines the evolution to the desired constraint manifold while simultaneously minimizing the energy. These results provide a first validation of the \texttt{CQITE} algorithm on a constrained optimization problem where the constraints compete directly with the Hamiltonian.

\subsection{Triangular-ladder spin model}
\label{sec:ladder}
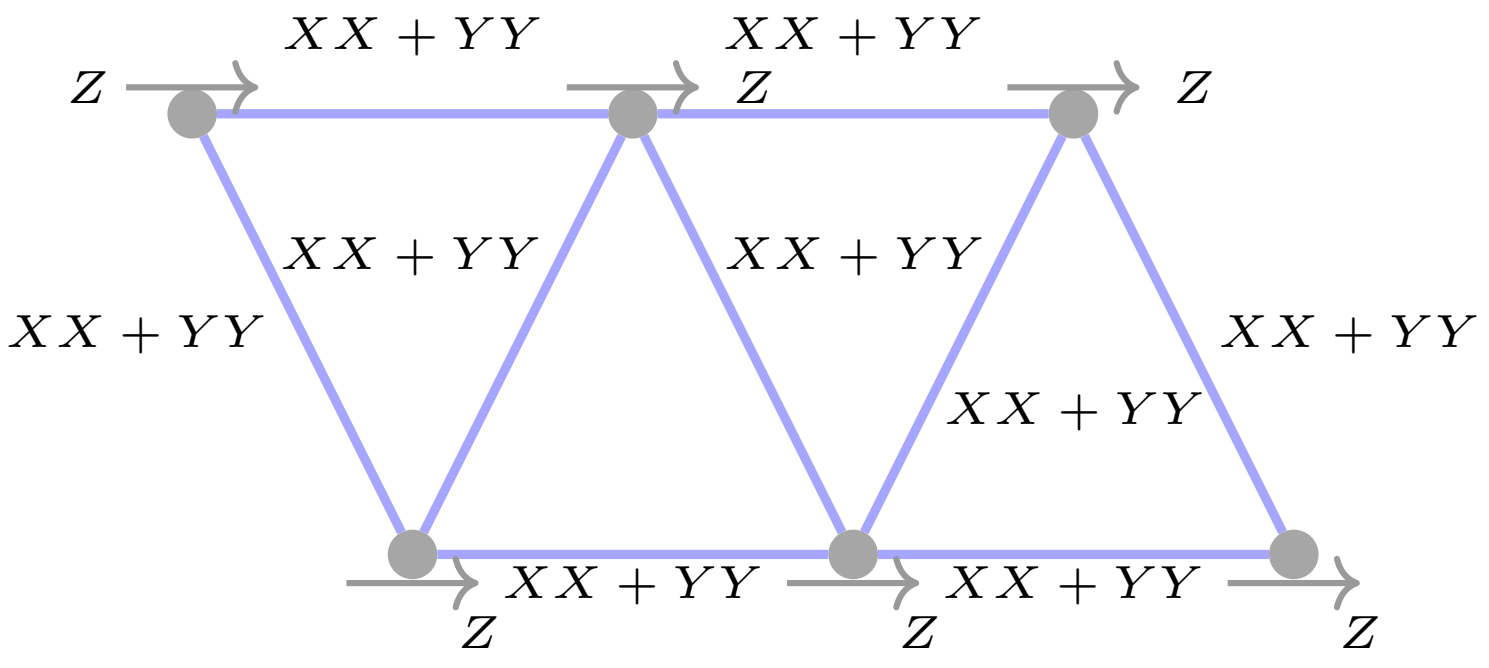
\begin{figure}[htb]
\centering
\resizebox{0.6\textwidth}{!}{
\begin{tikzpicture}[
spin/.style={
    circle,
    fill=black!35,
    draw=black!35,
    minimum size=6pt,
    inner sep=0pt
},
interaction/.style={
    very thick,
    draw=blue!35
},
ilabel/.style={
    font=\tiny
},
field/.style={
    ->,
    thick,
    draw=black!40
}
]


\node[spin] (s1) at (0,2) {};
\node[spin] (s2) at (2,2) {};
\node[spin] (s0) at (4,2) {};

\node[spin] (s3) at (1,0) {};
\node[spin] (s4) at (3,0) {};
\node[spin] (s5) at (5,0) {};


\draw[interaction]
(s1)--node[midway,above,ilabel, yshift=+3pt]
{$XX+YY$}(s2);

\draw[interaction]
(s1)--node[midway,left,ilabel]
{$XX+YY$}(s3);

\draw[interaction]
(s2)--node[midway,left,ilabel,yshift=10pt,xshift= 7pt]
{$XX+YY$}(s3);

\draw[interaction]
(s2)--node[midway,right,ilabel,yshift=10pt,xshift= -7pt]
{$XX+YY$}(s4);

\draw[interaction]
(s3)--node[midway,below,ilabel,yshift=3pt]
{$XX+YY$}(s4);

\draw[interaction]
(s2)--node[midway,above,ilabel,yshift=3pt]
{$XX+YY$}(s0);

\draw[interaction]
(s0)--node[midway,right,ilabel, yshift=-10pt, xshift=-7pt]
{$XX+YY$}(s4);

\draw[interaction]
(s4)--node[midway,below,ilabel,yshift=3pt]
{$XX+YY$}(s5);

\draw[interaction]
(s0)--node[midway,right,ilabel]
{$XX+YY$}(s5);


\draw[field]
($(s1)+(-0.3,+0.12)$)--++(0.6,0)
node[left,,ilabel,xshift=-15pt]{$Z$};

\draw[field]
($(s2)+(-0.3,+0.12)$)--++(0.6,0)
node[right,ilabel]{$Z$};

\draw[field]
($(s0)+(-0.3,+0.12)$)--++(0.6,0)
node[right,,ilabel]{$Z$};

\draw[field]
($(s3)+(-0.3,-0.13)$)--++(0.6,0)
node[right,below,ilabel]{$Z$};

\draw[field]
($(s4)+(-0.3,-0.13)$)--++(0.6,0)
node[right,below,ilabel]{$Z$};

\draw[field]
($(s5)+(-0.3,-0.13)$)--++(0.6,0)
node[right,below,ilabel]{$Z$};

\end{tikzpicture}
}
\caption{
Six-spin triangular-ladder spin model. Pairwise interactions
$\gamma_{uv}(X_u\otimes X_v+Y_u\otimes Y_v)$
are associated with the edges, while local magnetization constraints $Z$ are imposed on each vertex.
}
\hrulefill
\label{fig:ladder}
\end{figure}
\begin{figure}[htb]
    \centering
    \begin{subfigure}[b]{0.49\linewidth}
        \centering  \includegraphics[width=\linewidth]{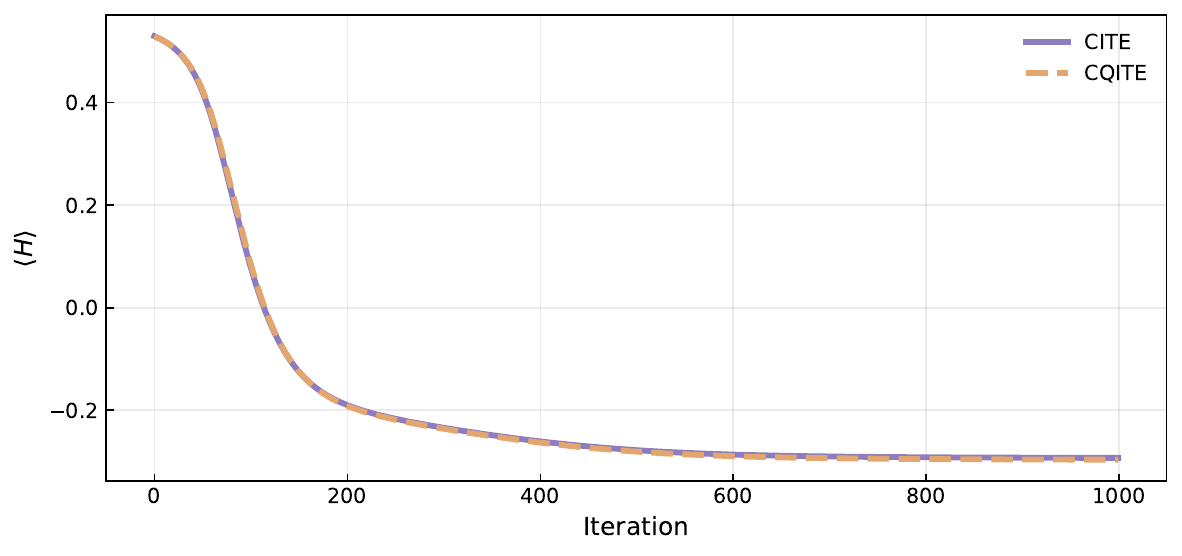}
        \caption{Energy evolution of \texttt{CQITE} and \texttt{CITE}.}
        \label{fig:zzxxy_exp}
    \end{subfigure}
    \hfill
    \begin{subfigure}[b]{0.49\linewidth}
        \centering
        \includegraphics[width=\linewidth]{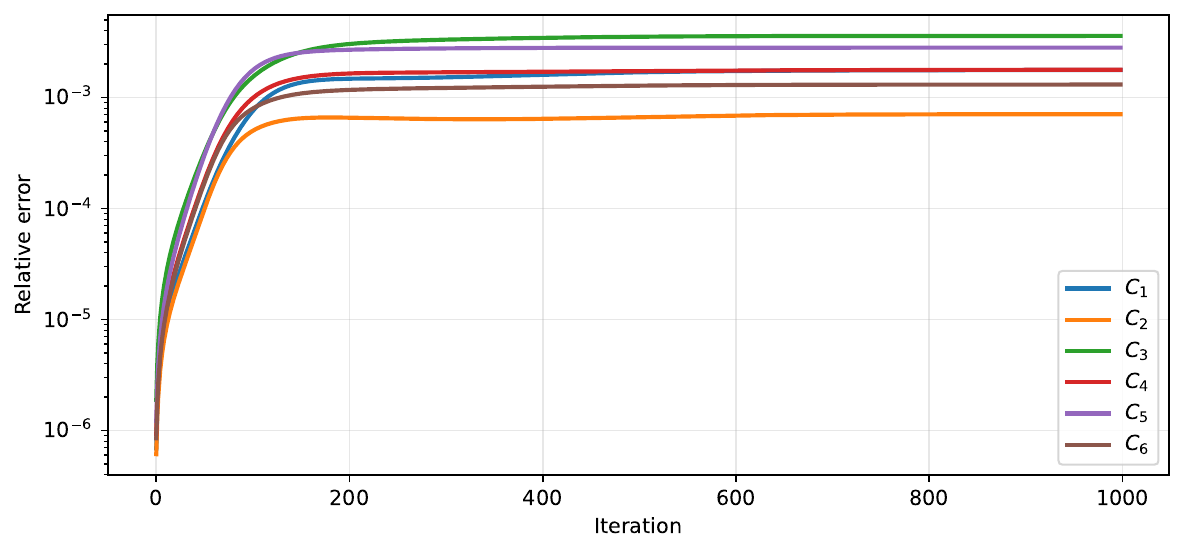}
    \caption{Constraint relative error.}
    \label{fig:zzxxy_con}
    \end{subfigure}
\caption{\texttt{CQITE} applied to the six-spin triangular-ladder model with local magnetization constraints. (a) Energy evolution obtained with homogeneous couplings $\gamma=0.2$, comparing \texttt{CQITE} with the exact \texttt{CITE}. (b) Evolution of the constrained expectation values $\langle\mathbf{Z}\rangle$ during the optimization for identical target values $\mathbf{b}=(.70, .75, .80, .85, .90, .95)$.
}
\hrulefill
\label{fig:zzxxy_results}
\end{figure}

As a second benchmark, we consider the six-spin triangular-ladder model shown in Fig.~\ref{fig:ladder}. Pairwise interactions $X_u X_v + Y_u Y_v$ are defined on the edges,
and local magnetization constraint $ \mathbf{Z} = (Z_1,\dots, Z_m)^{\mathsf T}$. The constrained optimization problem is

\begin{equation}
\begin{aligned}
\min_{|\varphi\rangle}\quad &
\sum_{(u,v)}
\gamma_{uv}
\langle
X_u X_v
+
Y_u Y_v
\rangle \\
\mathrm{s.t.}\quad &
\langle \mathbf Z\rangle=\mathbf b 
.
\end{aligned}
\label{eq:ising_constraint_problem_y}
\end{equation}

Fig.~\ref{fig:zzxxy_exp} compares the \texttt{CQITE} evolution with the constrained imaginary time evolution (\texttt{CITE}). The two trajectories are nearly indistinguishable throughout the optimization and converge to the same final energy. This agreement indicates that the Hamiltonian-generated operator pool accurately captures the tangent space explored by the constrained dynamics.



Finally, the constraint violations are shown in Fig.~\ref{fig:zzxxy_con}. Throughout the evolution, all six local constraints remain satisfied up to errors of order \(\sim\!10^{-3}\). Small residual deviations from the target constraint are nevertheless observed. Their origin is beyond the scope of the present work, although they do not appear to be associated with the Krylov operator-pool approximation, whose agreement with the full \texttt{CITE} dynamics remains excellent throughout the evolution.


\subsection{Two-particle fermionic triangle}
\label{sec:fermionic_triangle}

\begin{figure}[th]
    \centering
\includegraphics[width=0.7\linewidth]{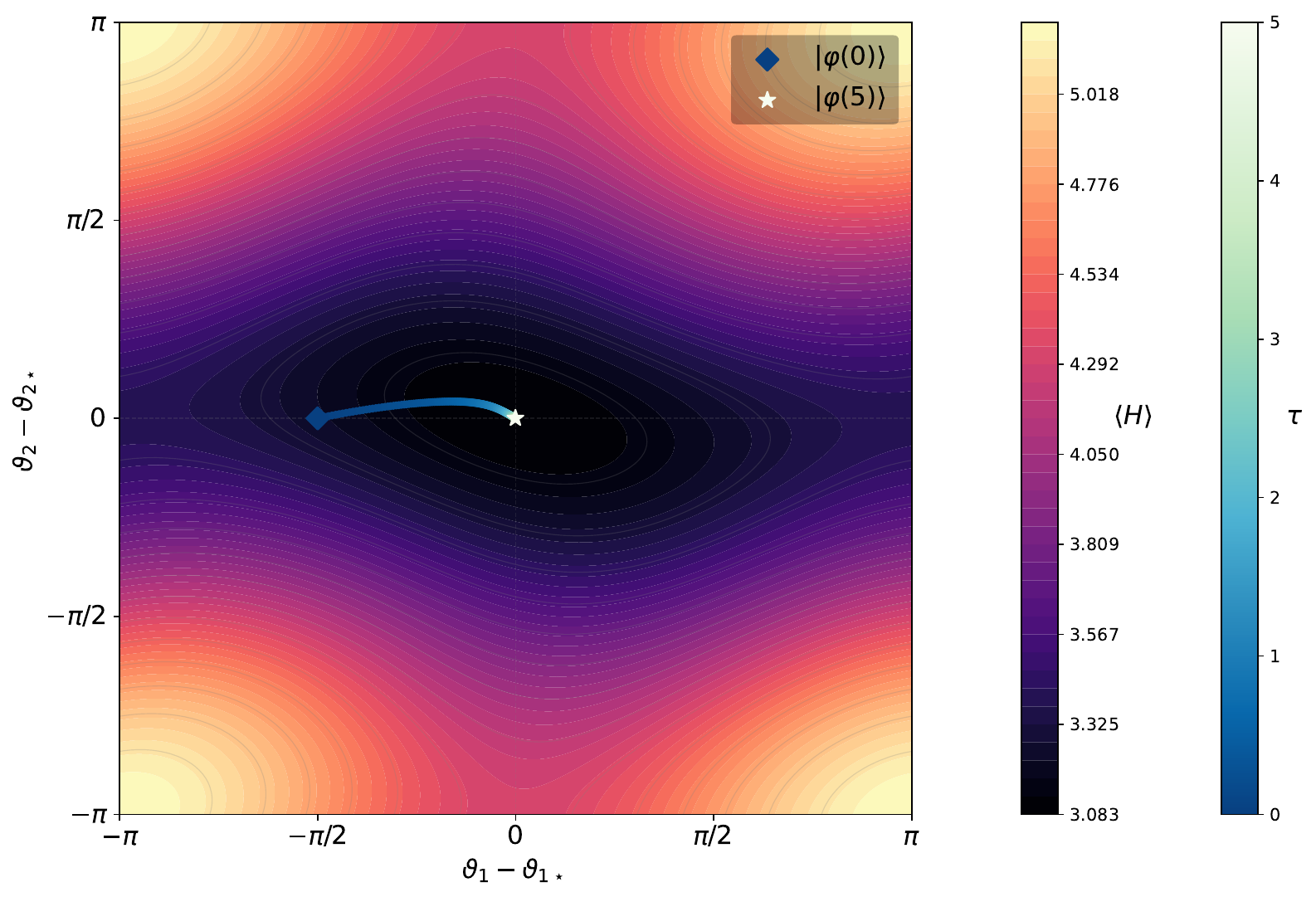}
\caption{\texttt{CQITE} evolution on the Penz--van Leeuwen constrained-search energy landscape in the phase coordinates $\vartheta_j - \vartheta_{j\star}$. The background shows the constrained energy range. The algorithm trajectory is represented by points coloured according to the imaginary time $\tau$, evolving from the initial state $|\varphi(0)\rangle$ to the final state $|\varphi(5)\rangle$. The origin corresponds to the constrained minimum.}
\hrulefill
\label{fig:fermionic_triangle_landscape}
\end{figure}

As a fermionic benchmark, we consider the two-particle triangle
graph studied by Penz and van Leeuwen~\cite{Penz2021} on which we consider two spin-less fermions. The system
is given by the Hamiltonian
\begin{equation}
    H = \sum_{p=1,q=1}^3 h_{pq} \, c_p^\dagger c_q^{\phantom{\dagger}}
\end{equation}
where 
\begin{equation}
h=
\begin{pmatrix}
\phantom{+}2&-1&-1\\
-1&\phantom{+}2&-1\\
-1&-1&\phantom{+}2
\end{pmatrix},
\end{equation}
which is the graph Laplacian of a triangular lattice with 3 sites.
We consider the $2$-particle sector. The constraints are given by the on-site particle density operators $\boldsymbol{\hat{\rho}} =(c_1^\dagger c_1^{\phantom{\dagger}},c_2^\dagger c_2^{\phantom{\dagger}},c_3^\dagger c_3^{\phantom{\dagger}})^{\mathsf T}$.
The goal is to minimise the expectation value of the Hamiltonian under specified values
$\rho_j = \langle \varphi | \hat{\rho}_j | \varphi \rangle$. This defines the so-called Levy constrained search density-functional
\begin{equation}
\begin{aligned}
F(\boldsymbol{\rho})=\min_{|\varphi\rangle}\quad
&\langle H\rangle\\
\text{s.t.}\quad
&\langle \boldsymbol{\hat{\rho}}\rangle=\boldsymbol{\rho}.
\end{aligned}
\end{equation}
where we minimize over all wave functions giving a pre-scribed density $\boldsymbol{\rho}=(\rho_1,\rho_2,\rho_3)^{\mathsf T}$. The densities lie on the plane $\rho_1+\rho_2+\rho_2=2$ in density space on which the conditions $0 \leq \rho_j \leq 1$ cut out a triangular density domain for the function $F ( \boldsymbol{\rho})$.

In the occupation basis $\left\{
\ket{110},
\ket{101},
\ket{011}
\right\}$, the wavefunction is given by
\begin{equation}
    | \varphi \rangle = c_1 \ket{110} + c_2 \ket{101} + c_3 \ket{011}
\end{equation}
the densities are given by
\begin{equation}
    \rho_1 = |c_1|^2 + |c_2|^2 \quad \rho_2 = |c_1|^2 + |c_3|^2 \quad \rho_3 = |c_2|^2 + |c_3|^2
\end{equation}
Since we can split off a global phase we can parametrize
\begin{equation}
     c_1=\sqrt{1-\rho_3} \, e^{i \vartheta_1} \quad c_2=\sqrt{1-\rho_2} \, e^{i \vartheta_2} \quad c_3=\sqrt{1-\rho_1}.
\end{equation}
and after a short calculation we find that
\begin{equation}
    F( \boldsymbol\rho ) = 4 + 2 \, \min_{\vartheta_1,\vartheta_2} \, T(\vartheta_1,\vartheta_2)
    \label{eq:func_phases}
\end{equation}
where 
\begin{equation}
     T(\vartheta_1, \vartheta_2) = \alpha_1 \, \cos \vartheta_1 - \alpha_2 \cos \vartheta_2 - 
     \alpha_3 \cos ( \vartheta_1 - \vartheta_2)
\end{equation}
where we defined
\begin{equation}
    \alpha_1 = \sqrt{ (1-\rho_1)(1 - \rho_3)} \quad \alpha_2 = \sqrt{ (1-\rho_1)(1 - \rho_2)} \quad 
    \alpha_3 = \sqrt{ (1-\rho_2)(1 - \rho_3)}
\end{equation}
The minimisation can be carried out analytically~\cite{Penz2021} and yields the exact expression:
\begin{equation}
    F (\boldsymbol\rho ) = \begin{cases} 3 & \| \boldsymbol\rho -\boldsymbol\rho_0 \|_2 \leq \frac{1}{\sqrt{6}} \\
    3 + ( 2 \max_k \sqrt{1-\rho_k} - \sum_{j=1}^3 \sqrt{1-\rho_j} )^2 & \| \boldsymbol\rho -\boldsymbol\rho_0 \|_2 > \frac{1}{\sqrt{6}} 
    \end{cases}
    \label{eq:triangle-functional}
\end{equation}
where $\boldsymbol\rho_0=(\frac23,\frac23,\frac23)^{\mathsf T}$ is the equidensity for equal occupation on all sites.
The densities satisfying $\| \boldsymbol\rho -\boldsymbol\rho_0 \|_2 \leq 1/\sqrt{6}$ are contained in a disc of radius $1/\sqrt{6}$ in the 
plane $\rho_1+\rho_2+\rho_3=2$ centered at $\boldsymbol\rho_0$. The densities in this disc correspond to degenerate
densities of the Hamiltonian in the absence of any external potential, while the remaining densities
in the density domain require a non-zero external potential to generate them as a ground state density.
The exception is the set of densities on the boundary of the triangular domain which correspond to full occupation $\rho_j=1$ on one of the sites and would require an infinite potential to generate (these are not $v$-representable in density-functional nomenclature), with the exception of the points where the degeneracy disc touches the boundary. 

Let us now consider the translation of the problem to qubits.
If we write the second quantized Hamiltonian in qubit form using a Jordan-Wigner transformstion we find
\begin{equation}
H = 4\mathds{1}
-\frac12
\left(
X_1 X_2+Y_1 Y_2
+X_2 X_3+Y_2 Y_3
+X_1 Z_2 X_3+Y_1 Z_2 Y_3
\right).
\label{eq:triangle_hamiltonian}
\end{equation}

Following the second-quantized formulation introduced in
Sec.~\ref{sec:sq}, the constraint operators are chosen
as the local occupation operators in the qubit representation,
\begin{equation}
\mathbf C =
\left(
Z_1,Z_2,Z_3
\right)^{\mathsf T}.
\end{equation}
The corresponding target satisfies
\begin{equation}
b_1 + b_2 + b_3
= -1.
\end{equation}
where the relations between $\rho_j$ and $b_j$ are given by $\mathbf{b} =1 - 2 \boldsymbol\rho$.
The initial state is selected within the fixed-density manifold as a non-stationary state of the Penz--van Leeuwen phase parametrization~\cite{Penz2021}.

To visualize the constrained-search landscape 
we consider the parametrization of Eq.\;\eqref{eq:func_phases} and display
the resulting constrained-search energy
landscape and the \texttt{CQITE} trajectory in
Fig.~\ref{fig:fermionic_triangle_landscape}.

The trajectory illustrates the relation between the \texttt{CQITE} dynamics and
the geometry of the constrained energy landscape. Starting from a
non-stationary point in the low-energy region, the evolution follows a
curved path towards the minimum, with its direction appearing to be
influenced by the surrounding higher-energy contours. Thus, the
phase-space representation provides an intuitive picture of how the
constrained imaginary-time dynamics navigate the local structure of the
energy surface. This complements the quantitative agreement between
\texttt{CQITE} and \texttt{CITE} observed in
the previous benchmarks.

\begin{figure}[ht]
\centering
\begin{subfigure}[b]{0.49\linewidth}
        \centering  
\includegraphics[width=0.95\linewidth]
{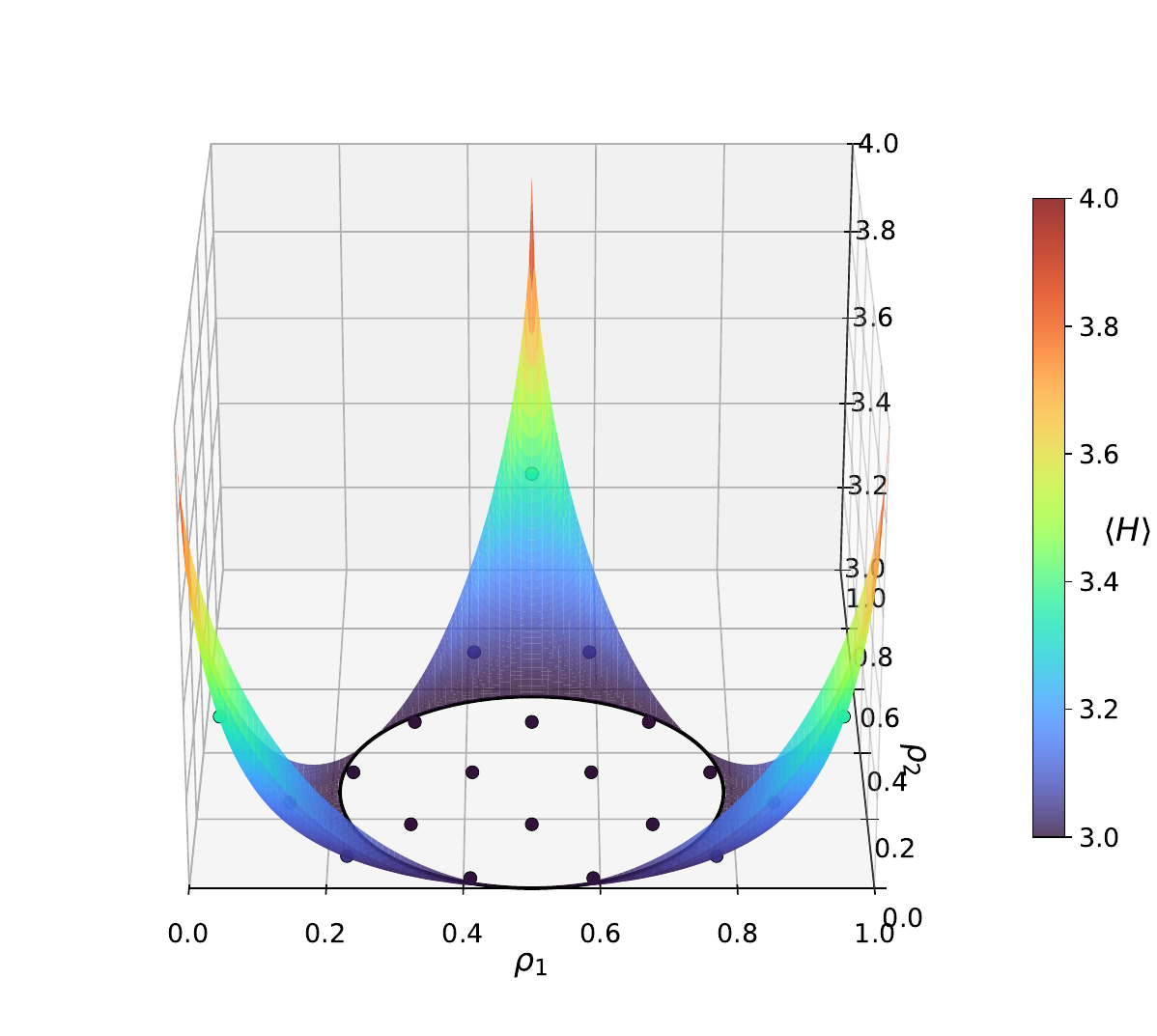}
\caption{Constrained energy and \texttt{CQITE} samples}
\end{subfigure}
\begin{subfigure}[b]{0.49\linewidth}
        \centering  
\includegraphics[width=0.92\linewidth]
{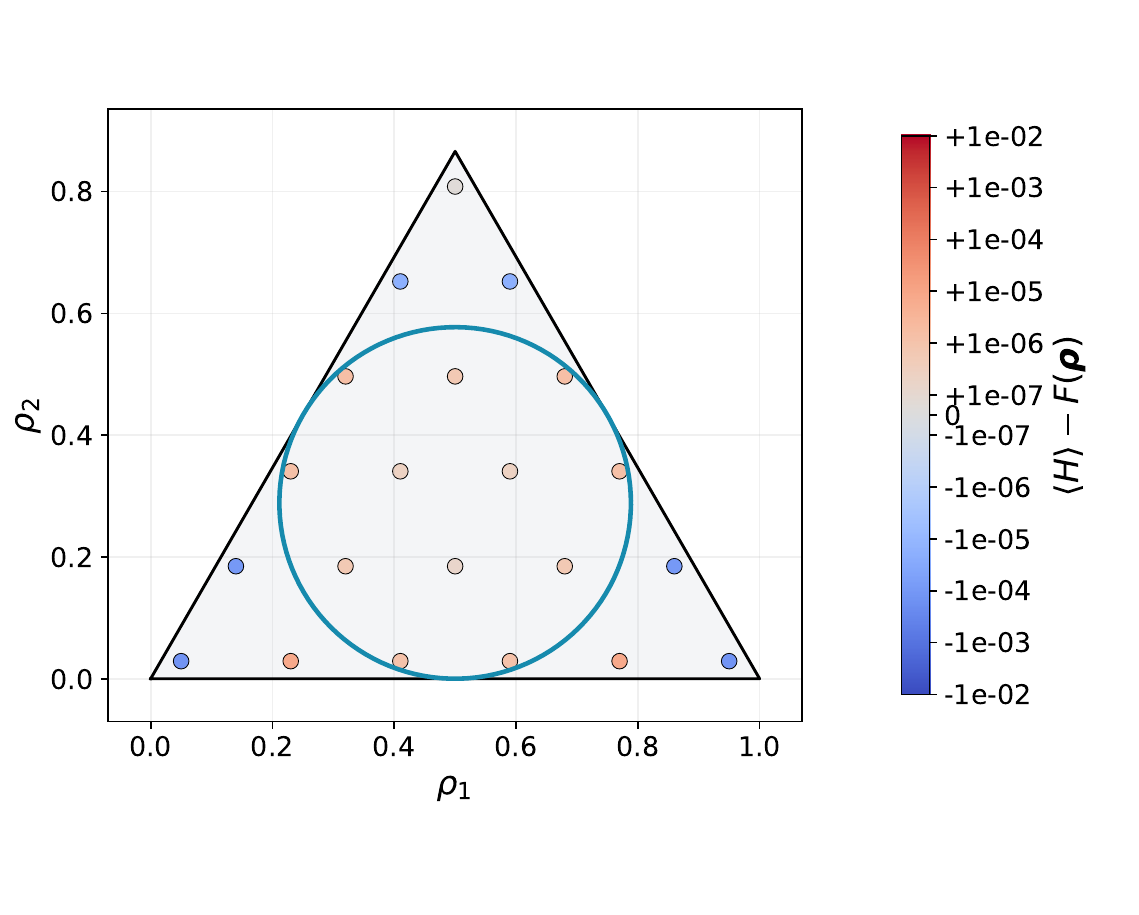}
\caption{Energy error across the density domain}
\end{subfigure}
    \caption{Exact constrained-search functional on the density domain of the three-site triangle graph, with converged \texttt{CQITE} energies at sampled target densities.}
    \label{fig:cqite-triangle}
\end{figure}

{\color{black}
We now consider the full constrained-search functional. 
For each target density $\boldsymbol{\rho}$, we initialize the \texttt{CQITE} evolution in the two-particle state
\begin{equation}
\lvert\varphi(0)\rangle
=
\sqrt{1-\rho_1}\,\lvert 011\rangle
+\sqrt{1-\rho_2}\,\lvert 101\rangle
+\mathrm{i}\sqrt{1-\rho_3}\,\lvert 110\rangle.
\end{equation}
The squared modulus of the amplitude of the basis state with site $j$
empty is $1-\rho_j$, and we choose the initial state to be complex since the densities in the degeneracy circle can only be reached by complex wavefunctions \cite{Penz2021}. The initial state has
$\langle \boldsymbol{\hat\rho} \rangle=\boldsymbol\rho$, or equivalently
$\langle \mathbf{Z}\rangle=1-2\boldsymbol\rho=\mathbf{b}$. It is normalized since $\sum_j \rho_j=2$.

Figure~\ref{fig:cqite-triangle}(a) compares the analytic reference over the full density domain with the energies obtained by \texttt{CQITE} at 21 target densities. These targets are obtained by contracting a uniform barycentric grid by a factor of \(0.9\) toward the domain centroid, preserving its triangular arrangement while placing all samples strictly inside the domain. The three outer branches rise away from the central region, where the functional is constant at its minimum, \(F(\boldsymbol\rho)=3\).

Figure~\ref{fig:cqite-triangle}(b) shows the signed energy difference \(\langle H \rangle-F(\boldsymbol\rho)\) at the same targets. The largest absolute energy difference is approximately \(1.4\times 10^{-4}\), with the largest deviations occurring at sampled densities in the outer regions.
}

The constraint potential $\mathbf{v}$ is the negative gradient of $F( \boldsymbol{\rho})$ \cite{Penz2021}:
\begin{equation}
    \mathbf{v} = - \nabla F ( \boldsymbol\rho)
\end{equation}
which gives $\mathbf{v}=0$ on the degeneracy disc and non-zero outside it. On the boundaries of the triangular domain, one of the density coordinates corresponds to full occupation, i.e. \(\rho_j=1\). We thus note from the terms \(\sqrt{1-\rho_j}\) in the expression for \(F(\boldsymbol{\rho})\) that, at these points, the gradient diverges as \((1-\rho_j)^{-1/2}\), making the constraint potential infinite.

\section{Summary and outlook}
In this work, we introduced constrained quantum imaginary-time evolution and derived a systematic Krylov hierarchy for constructing its operator pool. At fixed truncation order, the resulting pool grows polynomially with system size. Numerical results on spin and fermionic models show close agreement with constrained imaginary-time evolution and accurate constraint preservation.

Several directions merit further study. One is to investigate the Krylov construction for ground-state preparation with unconstrained Hamiltonians. It would also be useful to derive sharper truncation bounds that better account for the accuracy observed in our numerical experiments. Finally, benchmarking \texttt{CQITE} on quantum hardware could help assess the effects of finite sampling, noise, and circuit depth.

\section*{Acknowledgments}

All numerical experiments presented in this work were implemented and performed using \href{https://github.com/Qiskit/qiskit}{Qiskit}. The source code, together with the scripts required to reproduce the results reported in this paper, is publicly available on \href{https://github.com/danielecuomo/ConstraintQuantumImaginary}{GitHub}.

MS acknowledges support by the IRA Programme, project
no.~FENG.02.01-IP.05-0006/23, financed by the FENG program 2021--2027,
Priority FENG.02, Measure FENG.02.01, with the support of the FNP. LCB acknowledges support
from the Polish National Science Centre (NCN) through
the Sonata Bis (grant no. UMO-2024/54/E/ST2/00316).


\label{sec:references-manual}

\section*{Appendix}\label{sec:appendices-manual}
\appendix
\section{Constraint-based reference state}
\label{sec:init}

Let
\begin{equation}
\mathbf C
=
(C_1,\ldots,C_n)^{\mathsf T}\in\{X,Z\}^n.
\end{equation}
We restrict the constraints to local operators, as motivated by physically
relevant scenarios in which target constraint values are imposed through
local couplings to external fields or controls. Moreover, restricting the local constraints to $\{X,Z\}$ entails no loss
of generality, since any set of local Pauli constraints can be brought to this form by a local Clifford transformation, applied consistently to the
Hamiltonian, state, and constraint operators.

We introduce the Pauli vector
\[
\mathbf{Y}=(Y_1,\ldots,Y_n),
\]
which anticommutes with each constraint. For constraints acting on distinct qubits $[Y_j,C_k]=0$.
Starting from
\[
\ket{\mathbf{0}}=\ket{0}^{\otimes n},
\]
we define the constraint-dependent reference state
\[
\ket{\mathbf{C}}
=
\bigotimes_{j=1}^n
\begin{cases}
\ket{0}_j, & C_j=Z_j,\\[2pt]
\ket{+}_j, & C_j=X_j.
\end{cases}
\]
Equivalently,
\[
\ket{\mathbf{C}}
=
e^{+\ii\frac\pi4
\left(\mathbf{1}
-\langle\mathbf{0}|\mathbf{C}|\mathbf{0}\rangle\right)^{\mathsf T}\mathbf{Y}}
\ket{\mathbf{0}}.
\]

The target constraint values \(\mathbf b\) are then imposed by the rotation
\[
\ket{\varphi(0)}
=
e^{-\ii\frac{\arccos\mathbf b^{\mathsf T}}{2}
\mathbf{Y}}\ket{\mathbf C}.
\]
By construction,
\[
\langle\varphi(0)| \mathbf C|\varphi(0)\rangle
=
\mathbf b.
\]

\end{document}